\documentclass[prl,aps,showpacs,amsmath,amssymb,twocolumn]{revtex4}
\usepackage{epsfig,amsmath,amssymb,amsfonts}

\def\opone{\leavevmode\hbox{\small1\kern-3.8pt\normalsize1}}

\begin{document}

\title{Berry Phase in a Two-atom Jaynes-Cummings Model with Kerr Medium}

\author{Shen-Ping Bu, Guo-Feng Zhang, Jia Liu, Zi-Yu
Chen\footnote{Corresponding author.}\footnote{Email:
chenzy@buaa.edu.cn}}\affiliation{Department of Physics, School of
Science, BeiHang University, Xueyuan Road, Beijing 100083, P.R.
China}
\begin{abstract}
The Jaynes-Cummings model (JCM) is an very important model for
describing interaction between quantized electromagnetic fields and
atoms in cavity quantum electrodynamics (QED). This model is
generalized in many different direction since it predicts many novel
quantum effects that can be verified by modern physics experimental
technologies. In this paper, the Berry phase and entropy of the
ground state for arbitrary photon number $n$ of a two-atom
Jaynes-Cummings model with Kerr like medium are investigated. It is
found that there are some correspondence between their images,
especially the existence of a curve in the $\Delta-\varepsilon$
plane along which the energy, Berry phase and entropy all reach
their special values. So it is available for detecting entanglement
by applying Berry phase.
\end{abstract}

 \maketitle

\section{I. Introduction} Berry phase~\cite{Berry} or geometric phase, which does not have
classical correspondence, becomes a focus point in modern physics.
It describes a phase factor gained by the wavefunction after the
system undergoes an adiabatic and cyclic evolution, which reflects
the topological properties~\cite{Barry,Samuel} of the state space of
the system and has untrivial connections with the character of the
system~\cite{nonlocality}, especially with the
entanglement~\cite{Berry_entanglement1,Berry_entanglement2}.
Recently, the Berry phase was introduced into quantum computation to
construct a universal quantum logic gates that may be robust to
certain kinds of
errors~\cite{geometry-computation1,geometry-computation2,geometry-computation3,geometry-computation4,geometry-computation5}.

Cavity QED is an important solid-state system for implementing
quantum computation, and is studied extensively. In the theory of
cavity QED, the Jaynes-Cummings~\cite{jcm} model (JCM) is recognized
as the simplest and most effective model on the interaction between
radiation and matter, which can be solved exactly. As an important
theoretical model, JCM has led to many nontrivial predictions such
as collapse-revival pheno\-me\-non~\cite{collapse-revival},
squeezing~\cite{squeezing},
antibunching~\cite{antibunching1,antibunching2}, chaos~\cite{chaos},
and trapping states~\cite{trapping1,trapping2,trapping3}, etc.
Furthermore, despite the simplicity of JCM, it is of great
significance because recent technologies enabled scientists to
experimentally realize this rather idealized
mo\-del~\cite{experiments1,experiments2} and to verify some of the
theoretical predictions.

Stimulated by the success of the JCM, many people extends this model
to explore new quantum effects. One simple way of extending is
considering multiple atoms and multiple modes field instead of
single atom and single mode field~\cite{extensions1,extensions2}.
Another way is considering the interactions between field and medium
and fields itself, such as a cavity filled with Kerr medium.
Introducing of Kerr nonlinearity in the system Hamiltonian will
cause various nonlinear effects, so it attracts much attentions of
scientists~\cite{nonlinear1,kerr1,kerr2,kerr4,kerr3}. One of the
many applications of these nonlinear effects is to produce entangled
states~\cite{kerr_entanglement}, which is of extensively
applications in quantum information, especially in quantum
communication.

In this paper, we try to investigate a two-atom Jaynes-Cum\-mings
model in Kerr medium. At first, we calculate the eigenvalues and
eigenstates of the system. Then we evaluate the Berry phase of
ground state for arbitrary photon number $n$ in terms of the
introduction of the phase shift operator, and for comparing the
phase with the entanglement we compute the von Neumann entropy as a
measurement of entanglement. After these tedious computation, we
compare the ground state energy, Berry phase and entropy, and find
that there are tight connections between them.

\section{II. Hamiltonian and Ground State Energy}

The Hamiltonian of the system in the rotating wave approximation can
be written as (assuming~ $\hbar =1$ )
\begin{eqnarray}
\label{eq:hamilton}
 H&=&\omega_{f}a^{\dagger}a
+\frac{\omega_{0}}{2}\sum_{j=1}^{2}\sigma_{z}^{j}+\varepsilon\sum_{j=1}^{2}\left(a\sigma_{+}^j+a^{\dagger}\sigma_{-}^j\right)
\nonumber\\&&+\chi\left(a^{\dagger}aa^{\dagger}a\right)
\end{eqnarray}
where $a^{\dagger}$ and $a$ denote the creation and annihilation
operators of the single mode field, $\omega_{f}$ is the transition
frequency of the field, $\omega_{0}$ is the atomic transition
frequency, $\varepsilon$ is the coupling constant between these two
atoms and field, $\chi$ represents the coupling of the fields
induced by the Kerr medium. $\sigma_{z}=\left|e\rangle_{j}\langle
g\right|-\left|g\rangle_{j}\langle e\right|$,
$\sigma_{+}^{j}=\left|e\rangle_{j}\langle g\right|$,
$\sigma_{-}^{j}=\left|g\rangle_{j}\langle e\right|$, with
$\left|e\right\rangle_{j}$ and $\left|g\right\rangle_{j}$ being the
excited and ground states of $j$ th atom, $j=1,2$. By the way, there
exists a conserved quantity $K$ for above Hamiltonian, which is
\begin{equation}
\label{eq:K}
K=a^{\dagger}a+1+\frac{\sigma_{z}^{1}+\sigma_{z}^{2}}{2}.
\end{equation}
The basis of the subspace $\left(K=n+2\right)$ is
\begin{eqnarray*}
\label{eq:basis}
\left|n,e_{1},e_{2}\right\rangle&,&\left|n+1,e_{1},g_{2}\right\rangle,\nonumber \\
\left|n+1,g_{1},e_{2}\right\rangle&,&\left|n+2,g_{1},g_{2}\right\rangle.
\end{eqnarray*}
And in that basis, the Hamiltonian is written as (in an appropriate
interaction picture)
\begin{equation}
\label{eq:matrixhamilton} H=\left(
\begin{array}{cccc}
\Delta-\chi(2n+2)& \varepsilon\sqrt{n+1}& \varepsilon\sqrt{n+1}& 0\\
\varepsilon\sqrt{n+1}& -\chi& 0& \varepsilon\sqrt{n+2}\\
\varepsilon\sqrt{n+1}& 0& -\chi& \varepsilon\sqrt{n+2}\\
0& \varepsilon\sqrt{n+2}& \varepsilon\sqrt{n+2}& \Delta-\chi(2n+2)
\end{array}
\right).
\end{equation}
where $\Delta=\omega_{0}-\omega_{f}$ is the detuning of the cavity
field. The four eigenvalues $\lambda_{j} (j=1,2,3,4)$ and
corresponding eigenstates $\left|\psi_{j}\right\rangle$ have been
calculated analytically. However, it is useless to present their
complicated formulas here, but $\left|\psi_{j}\right\rangle$ can be
written simply as follows:
\begin{eqnarray}
\label{eq:eigenstates} \left|\psi_{j}\right\rangle&
=&c_{j}^{1}\left|n,e_{1},e_{2}\right\rangle+c_{j}^{2}\left|n+1,e_{1},g_{2}\right\rangle\nonumber\\&&+c_{j}^{3}\left|n+1,g_{1},e_{2}\right\rangle
+c_{j}^{4}\left|n+2,g_{1},g_{2}\right\rangle
\end{eqnarray}
When $n=0$, the ground state energy, ie. the lowest eigenvalue of
the Hamiltonian for $n=0$ , as a function of detuning $\Delta$ and
coupling constant $\varepsilon$, is shown in Fig.~1. We find in the
figure that when $\varepsilon$ approaches $0$ there exist two
discontinuity points in the derivative of the energy, and the image
of the function are symmetry against the line $\Delta=2$ to some
extent. As we will see, these two points will be singularities of
Berry phase, and the symmetry of the energy function will also be
inherited by the Berry phase.
\begin{figure}
\begin{center}
\epsfig{figure=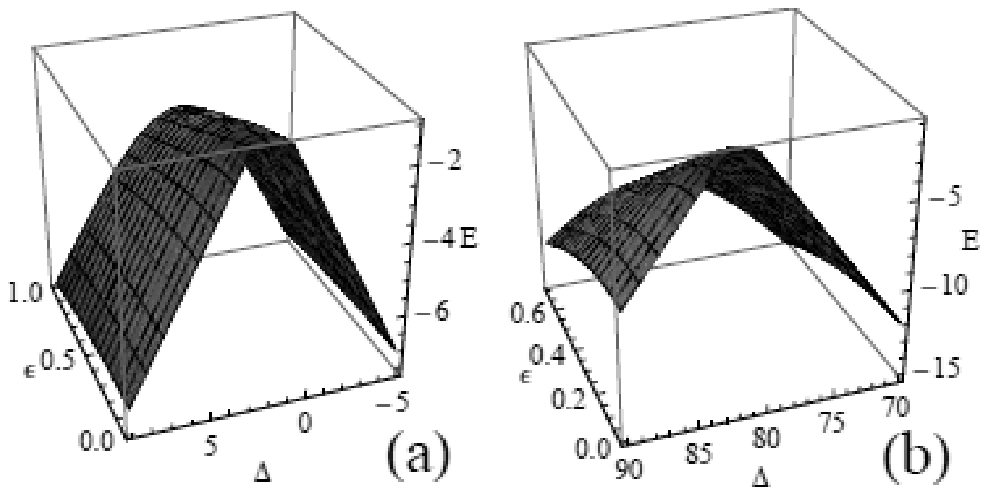,width=0.48\textwidth}
\end{center}
\caption{(Color online) Ground state energy versus detuning $\Delta$
and coupling constant $\varepsilon$ , where $\Delta$ and
$\varepsilon$ are measured with unit $\chi$. Part (a) and (b) are
respectively correspond to the case $n=0$ and $n=40$ .}
\end{figure}

\section{III. Berry Phase and Entropy}

Obviously, the whole system are quantized, to study the geometric
properties of this system we resort to the method of
\cite{computeberryphase} to evaluate the Berry phase of the system
by introducing a phase shift operator:
\begin{equation}
R\left(t\right)=e^{-i\varphi\left(t\right)a^{\dagger}a}
\end{equation}
where $\varphi\left(t\right)$ is changed from $0$ to $2\pi$
adiabatically. Then the time independent eigen equation of the
system:
$H\left|\psi_{j}\right\rangle=\lambda_{j}\left|\psi_{j}\right\rangle$
is changed into
$H'\left|\psi'_{j}\right\rangle=\lambda'_{j}\left|\psi'_{j}\right\rangle$,
with
$H'=R\left(t\right)HR^{\dagger}\left(t\right)-iR\left(t\right)dR^{\dagger}\left(t\right)/dt$
and
$\left|\psi'_{j}\right\rangle=R\left(t\right)\left|\psi_{j}\right\rangle$.
Hence the Berry phase can be evaluated according to the standard
method as follows:
\begin{eqnarray}
\label{eq:berry phase}
\gamma_{j}&=&i\int_{0}^{2\pi}d\varphi\langle\psi'_{j}\left|\frac{d}{d\varphi}\right|\psi'_{j}\rangle\nonumber\\
&=&i\int_{0}^{2\pi}d\varphi\langle\psi_{j}\left|R^{\dagger}\left(t\right)\frac{d}{d\varphi}R\left(t\right)\right|\psi_{j}\rangle
\end{eqnarray}
For our model, the Berry phase is given as
\begin{eqnarray}
\label{eq:computation fomulus}
\gamma_{j}&=&2\pi\left[n\left|c_{j}^{1}\right|^{2}+\left(n+1\right)\left(\left|c_{j}^{2}\right|^{2}+\left|c_{j}^{3}\right|^{2}\right)\right. \nonumber\\
&&{\left. +\left(n+2\right)\left|c_{j}^{4}\right|^{2}\right]}
\end{eqnarray}
Apparently, $c_{j}^{i}(i,j=1,2,3,4)$ are functions of detuning
$\Delta$ and coupling constant $\varepsilon$. So the Berry phase of
the ground state can be controlled by $\Delta$ and $\varepsilon$.
Fig.~2 shows its image of the case $n=0$ .
\begin{figure}
\begin{center}
\epsfig{figure=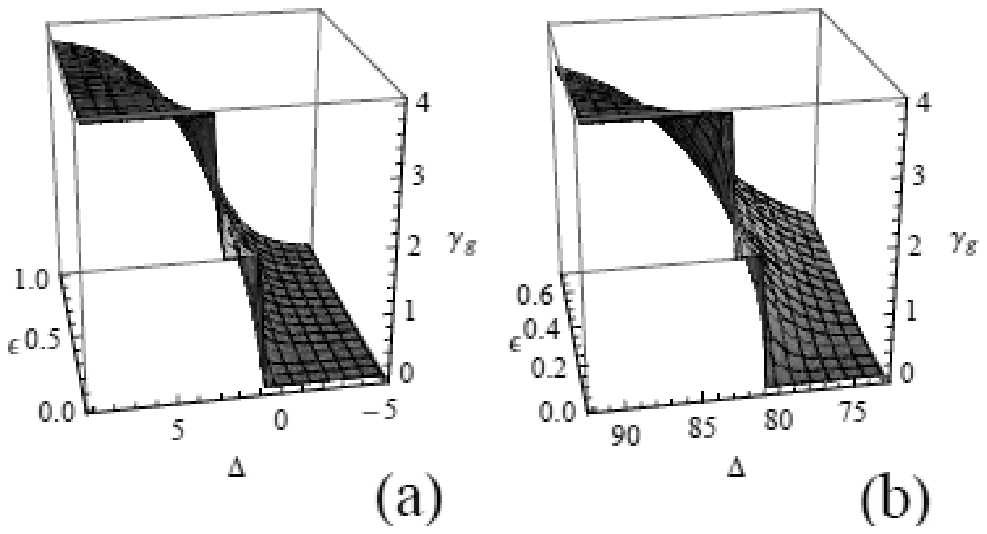,width=0.48\textwidth}
\end{center}
\caption{(Color online) Ground state Berry phase versus detuning
$\Delta$ and coupling constant $\varepsilon$ where $\Delta$ and
$\varepsilon$ are measured with unit $\chi$, and the unit of the
vertical axis is $\pi$. Part (a) and (b) are respectively correspond
to the case $n=0$ and $n=40$ .}
\end{figure}

Just as we have mentioned before, there are two singularities when
$\varepsilon$ approaches $0$ for the Berry phase, and the image is
centrosymmetric to some extent against the intersection curve of the
Berry phase image and $2\pi$ plane where Berry phase identical
equals to $2\pi$, which is adjacent to the plane $\Delta=2$ . This
result is similar to that of \cite{Tavis-Cummings}. In the
article~\cite{Tavis-Cummings}, the authors calculate the Berry phase
of ground state of Tavis-Cummings Model, and it is also found that
there are correspondence between the singularities of Berry phase
and ground state energy as well as symmetry.

To compare the Berry phase with the entanglement of the system, we
calculate the entropy, which can be used to measure the
entanglement, using following definition:
\begin{equation}
\label{eq:entropy}
S\left(\rho_{a}\right)=-\mbox{tr}\left[\rho_{a}\log_{2}\left(\rho_{a}\right)\right]
\end{equation}
where $\rho_{a}=\mbox{tr}^{f}\left(\rho_{af}\right)$ is the reduced
density operator of $\rho_{af}$, and $\rho_{af}$ represents the
density operator of the system. Generally, when the system is in a
pure state $\left|\psi\right\rangle$, its density operator
$\rho_{af}=\left|\psi\rangle\langle\psi\right|$.

According to our computations, we present the figure of entropy as a
function of detuning $\Delta$ and coupling constant $\varepsilon$
when $n=0$ in Fig.~3. Apparently as the figures show, there are the
same two points and symmetry correspond to that of the Berry phase
and energy.
\begin{figure}
\begin{center}
\epsfig{figure=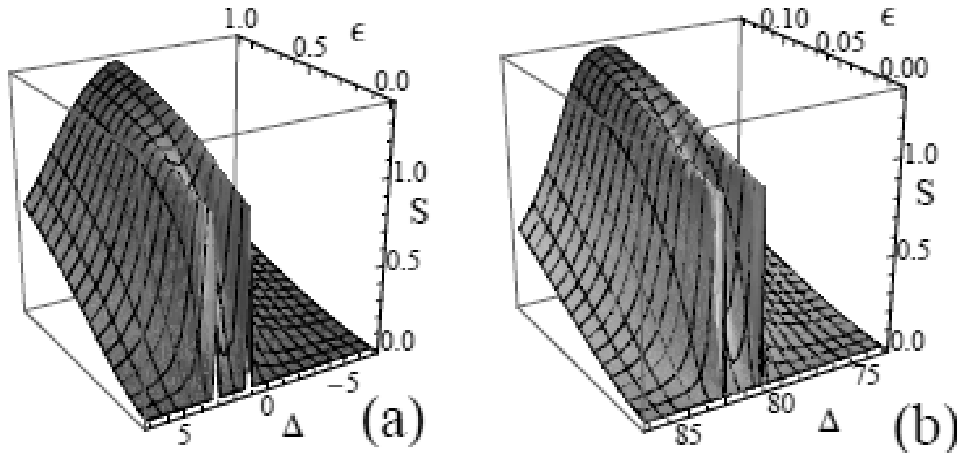,width=0.48\textwidth}
\end{center}
\caption{(Color online) Ground state entropy versus detuning
$\Delta$ and coupling constant $\varepsilon$ , where $\Delta$ and
$\varepsilon$ are measured with unit $\chi$. Part (a) and (b) are
respectively correspond to the case $n=0$ and $n=40$ .}
\end{figure}

According to  the image and our calculations, we find that for each
different value of $\varepsilon$, there exists a maximum value for
the energy of the ground state when $\Delta$ satisfies following
equation:
\begin{equation}
\label{eq:0_maximumline}
\Delta=\frac{1}{2}+\sqrt{2}-\frac{1}{2}\sqrt{17-12\sqrt{2}+\left(12-8\sqrt{2}\right)\varepsilon^{2}}
\left(\varepsilon\neq0\right).
\end{equation}
And to our surprise, at these points where $\Delta$ and
$\varepsilon$ satisfy above equation, the Berry phase is right
$2\pi$ and the entropy of the system reached its relative minimum
value when $\varepsilon$ is near $0$. The equation determines a
curve in the $\Delta-\varepsilon$  plane and because this curve
reflects the main character of ground state, we call it the
characteristic curve of the ground state. The existence of the
characteristic curves proved the tight connections between energy,
Berry phase and entanglement.
\begin{figure}
\begin{center}
\epsfig{figure=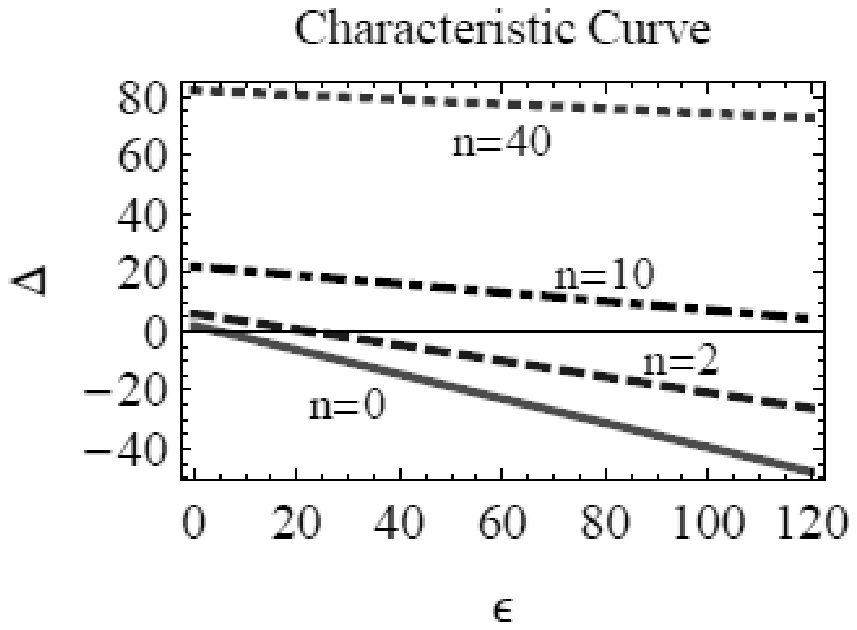,width=0.48\textwidth}
\end{center}
\caption{(Color online) The characteristic curves of the case $n=0$
, $n=2$ , $n=10$ , and $n=40$ . It is clearly showed that the move
of the curve with the value of $n$ increasing. }
\end{figure}

We also considered the case $n\neq0$, and find that the images of
ground state energy, Berry phase and entropy versus $\Delta$ and
$\varepsilon$ are similar to the case $n=0$, such as the symmetry
against a line $\left(\Delta=2n+2\right)$ to some extent, and the
correspondence of singularities. To illustrate this, we represent
the images of ground state energy, Berry phase and entropy when
$n=40$ in part (b)'s of Fig.~1, Fig.~2 and Fig.~3 respectively.
Obviously, the main difference between them is that the move of the
characteristic curve in the $\Delta-\varepsilon$ plane and its
equation reads $\left(\varepsilon\neq0\right)$  :
\begin{eqnarray}
\label{eq:n_maximumline} \Delta & = &
\frac{1}{2}\left(2n+1\right)+\mbox{A}\nonumber\\
&&-\frac{1}{2}\sqrt{1-4\left(2n+3\right)\mbox{A}+8\mbox{A}^{2}+\left(12+8n-8\mbox{A}\right)\varepsilon^{2}}\nonumber\\
\mbox{A} & = & \sqrt{n^{2}+3n+2} .
\end{eqnarray}
Fig.~4 shows the characteristic curves for different values of $n$.
We think this result maybe owe to the fact that the Berry phase and
entropy are all functions of the ground state energy.

\section{IV. Conclusion}
In conclusion, we calculated the Berry phase and entropy of a
two-atom Jaynes-Cummings model with Kerr medium, and found that
there are correspondences between their singularities and symmetry.
Especially, there exist a class of curves in the
$\Delta-\varepsilon$ plane, along which the Berry phase and entropy
all reach their special values like $2\pi$ for Berry phase. These
results reflect the tight relations between the Berry phase and
entanglement of the system, and maybe it is caused by the fact that
they are all functions of energy. Some physicists are trying to
measure entanglement using Berry phase, and our results may be
useful to them.

\section{acknowledgements}
This work was supported by the National Natural Science Foundation
of China (Grant No. 10604053) and the Beihang Lantian Project;G. F.
Zhang also acknowledges the support of the National Natural Science
Foundation of China (Grant No. 10874013)

\end{document}